\documentclass[aps,prb,twocolumn,floats,showpacs,superscriptaddress,longbibliography]{revtex4-1}

\usepackage{bm}
\usepackage{graphicx}
\usepackage{epstopdf}
\usepackage{amsmath}
\usepackage{times}

\newcommand{\vecb}[1]{\mathbf{#1}}

\def \FUW{Institute of Experimental Physics, Faculty of Physics, University of Warsaw, Pasteura 5, 02-093 Warsaw, Poland}
\def \GRENOBLE{Laboratoire National des Champs Magn\'etiques Intenses CNRS-UGA-UPS-INSA-EMFL, 30942 Grenoble, France}

\begin{document}

\title{Direct determination of zero-field splitting for single Co$^{2+}$ ion\\embedded in a CdTe/ZnTe quantum dot} 
\author{J. \surname{Kobak}}\email{Jakub.Kobak@fuw.edu.pl}
\affiliation{\FUW}
\author{A. \surname{Bogucki}}\affiliation{\FUW}
\author{T. \surname{Smole\'nski}}\affiliation{\FUW}
\author{M. \surname{Papaj}}\affiliation{\FUW}
\author{M. \surname{Koperski}}\affiliation{\FUW}\affiliation{\GRENOBLE}
\author{M. \surname{Potemski}}\affiliation{\GRENOBLE}
\author{P. \surname{Kossacki}}\affiliation{\FUW}
\author{A. \surname{Golnik}}\affiliation{\FUW}
\author{W. \surname{Pacuski}}\affiliation{\FUW}

\begin{abstract} 
When Co$^{2+}$ impurity is embedded in semiconductor structure, crystal strain strongly influences zero-filed splitting between Co$^{2+}$ states with spin projection $S_z = \pm 3/2$ and $S_z = \pm 1/2$. Experimental evidences of this effect have been given in previous studies, however direct measurement of the strain induced zero-field splitting has been inaccessible so far. Here this splitting is determined thanks to magneto-optical studies of individual Co$^{2+}$ ion in epitaxial CdTe quantum dot in ZnTe barrier. Using partially allowed optical transitions we measure strain induced zero-field splitting of Co$^{2+}$ ion directly on excitonic photoluminescence spectrum. Moreover, by observation of anticrossing of $S_z = + 3/2$ and $S_z = - 1/2$ Co$^{2+}$ spin states in magnetic field, we determine  axial and in-plane components of crystal field acting on Co$^{2+}$. Proposed technique can be applied for optical determination of zero-field splitting of other transition metal ions in quantum dots.

\end{abstract}


\maketitle

\section{Introduction}
The first observation of excitonic emission from quantum dot (QD) with a single Co$^{2+}$ ion\cite{kobak-nature-2014} led to the discovery that the quantum dot containing single magnetic ion exhibits efficient radiative excitonic recombination, even in case when recombination energy is higher than intraionic transition of magnetic ion.
Therefore optical manipulation of single magnetic ion spin is possible not only in classical systems with single Mn$^{2+}$ in QDs with relatively low energy gap CdTe/ZnTe\cite{Besombes_2004_PRL,Besombes_2005_PRB,LeGall_2009_PRL,Goryca_2009_PRL2,Goryca_2010_PRB} and InAs/GaAs\cite{Kudelski_2007_PRL,Krebs_2009_PRB, Baudin_2011_PRL}, but also in QDs  with higher energy gap such as CdSe/ZnSe\cite{kobak-nature-2014,Smolenski_2015_PRB,Fainblat_NL_2016} or in QD systems doped with impurities considered previously as killers of photoluminescence: Co$^{2+}$,\cite{kobak-nature-2014} Fe$^{2+}$,\cite{Smolenski_2016_Nature_Comm}  Cr$^{2+}$.\cite{Lafuente_Sampietro_2016_PRB,Lafuente_APL_2016} For all the systems studied so far it has been found that ground state of the magnetic ion is significantly influenced by the local strain present in a quantum dot. Such effect is of particular importance for the future applications in solotronics (optoelectronics based on single dopants\cite{Koenraad_2011_Nature_Mat}) and spintronics based on magnetic QDs.\cite{Beaulac_Science_2009,Whitaker_NL_2011,Pandey_NN_2012,Qu_PRB_2015,Barman_PRB_2015,Loureco_JPCC_2015,Rice_NN_2016,Nistor_JAC_2016,Moldoveanu_PRB_2016,Balanta_PRB_2016,Muckel_AN_2016} \mbox{Effect} of strain may be demonstrated as the beating in coherent precession of single magnetic ion spin,\cite{Goryca_PRL_2014} as changes of occupation and energy of levels corresponding to various projection of ion spin,\cite{Kudelski_2007_PRL,kobak-nature-2014,Lafuente_Sampietro_2016_PRB}, or even as the change of the character of Fe$^{2+}$ ion ground state, from nonmagnetic in bulk, to the magnetic in QD.\cite{Smolenski_2016_Nature_Comm} However so far only indirect method were used for estimation of strain induced zero-field splitting of  single magnetic ions in QDs: from depolarization efficiency of optically oriented ion,\cite{LeGall_2009_PRL} from PL peaks line-shape\cite{Besombes_PRB_2014}, from variation of coherent oscillation amplitude,\cite{Goryca_PRL_2014,Lafuente_Sampietro_2015_PRB} and from relative PL peaks intensity of lines corresponding to various spin-projection.\cite{Kudelski_2007_PRL,kobak-nature-2014,Lafuente_Sampietro_2016_PRB,Smolenski_2016_Nature_Comm} Here we present a method for measurement of strain induced zero-field splitting of magnetic ion directly by observation of corresponding splitting in exciton photoluminescence spectrum. We apply this method for precise determination of zero-field splitting of Co$^{2+}$ in CdTe/ZnTe QD.

The paper is organized in the following way: first we give experimental details and we recall typical spectrum of the neutral exciton in a QD with a single Co$^{2+}$ ion. Next we introduce partially allowed transitions which are used  to determine zero-field splitting of a single Co$^{2+}$ ion in a QD. Finally we present magneto-optical effects of anticrossing of Co$^{2+}$ spin states. From this anticrossing we determine Hamiltonian parameters $D$ and $E$ representing axial and in-plane components of the crystal field.

\section{Samples and experimental setup}

We have studied $2$ similar structures with ZnTe barriers and CdTe QDs doped with Co$^{2+}$ ions. The growth of investigated structures was performed by molecular beam epitaxy on GaAs (100) oriented substrate. Formation of the QDs was induced by amorphous Te desorption method.\cite{tinjod-apl-2003} During the CdTe deposition process we introduced low level delta--doping with Co$^{2+}$ ions using high temperature effusion cell working at $1200^{\circ}$C. Since molecular fluxes of cobalt used during the growth of QDs are smaller than the accuracy of the vacuum gauge, for both investigated structures we fabricated the reference Zn$_{1-x}$Co$_x$Te layers grown under the same Co$^{2+}$ flux. Such layers of diluted magnetic semiconductor allowed us to determine precisely Co$^{2+}$ concentration $x$ by measuring the giant Zeeman effect (more details in \onlinecite{papaj-JCG-2014,papaj-appa-2012}). The final Co$^{2+}$ concentration in reference samples was found to be equal about $0.3$\%. The same concentration was therefore expected in thin Cd$_{1-x}$Co$_x$Te layer before transformation into QDs. 

Maximal compressive strain in our QDs is related to lattice mismatch of about 6\% defined by lattice constance of CdTe (a = 6.48 {\AA}) and ZnTe (a = 6.10 {\AA}). In practice, strain can be lower due to intermixing of Zn and Cd between QDs and barrier or due to partial relaxation of 2 monolayer thick CdTe layer during formation of QDs.

For microphotoluminesce experiments the samples were immersed in liquid helium inside a magneto-optical bath cryostat with magnetic field of up to 10 T. The measurements were performed at the temperature of about 1.5 K using a high resolution reflective microscope, which results in about 0.5 $\mu$m diameter of laser spot. Such experimental setup allowed us to study optical properties of well-separated emission lines in magnetic fields with a polarization resolution. Complementary studies requiring a high magnetic field were performed in Grenoble High Magnetic Field Laboratory, where samples were immersed in helium gas at T = 10~K inside a 20~MW resistive magnet producing magnetic field of up to 28~T.

\section{Results}

\subsection{Typical photoluminescence spectrum of a QD with a single Co$^{2+}$ ion}

\begin{figure}
\includegraphics[width=0.7\linewidth]{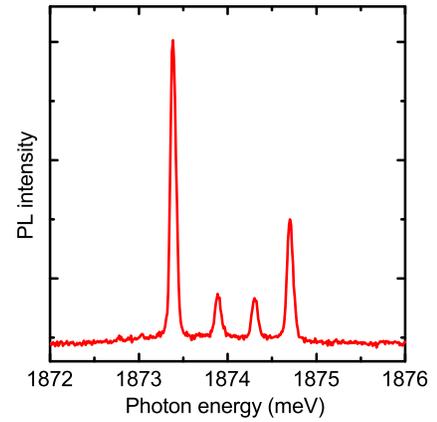}
\caption{(Color online) Typical photoluminescence spectrum of a neutral exciton in a CdTe/ZnTe QD with a single Co$^{2+}$ ion at T = 3~K. This Co$^{2+}$ ion has a fundamental state with spin projection $\pm3/2$ and as a consequence outer lines related to such a spin states are much more pronounced than inner emission lines related to $\pm1/2$ states.}
\label{Fig4lines}
\end{figure}

\begin{figure}
\includegraphics[width=1\linewidth]{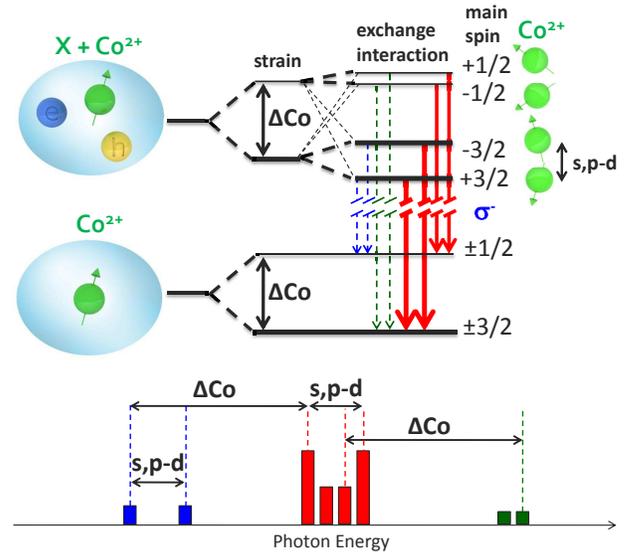}
\caption{(Color online) A schematic diagram of zero-field optical transitions between energy levels of singly Co$^{2+}$-doped quantum dot with and without neutral exciton. Numbers indicate main Co$^{2+}$ spin wavefunction. Red solid arrows denote the transitions for which Co$^{2+}$ spin is conserved - the only allowed optical transitions of bright exciton if Co$^{2+}$ states are pure. Blue and green dashed arrows denote transitions partially allowed due to Co$^{2+}$ spin state mixing. Energy difference between the strongest emission lines in spectra (red outer lines) and weak, partially allowed lines at lower energy (blue lines) are equal exactly to cobalt zero-field splitting ($\Delta_{Co}$).}
\label{NaprezeneCo_TransitionScheme}
\end{figure}

Fig.~\ref{Fig4lines} presents photoluminescence spectrum of a bright neutral exciton in a CdTe/ZnTe QD with a single Co$^{2+}$ ion in zero magnetic field. Emission is split due to $s,p$-$d$ exchange interaction between exciton and magnetic ion. Four observed lines result from four possible spin projection of Co$^{2+}$ ion with spin $3/2$ onto the quantization axis given by heavy-hole exciton in QD.\cite{kobak-nature-2014} More precisely, in excited state formed by exciton and Co$^{2+}$ ion, for each projection of excitonic spin ($\pm1$,~corresponding to $\sigma^+$, $\sigma^-$ circular polarisations) there are four energy levels and in the final state after exciton recombination there are four quantum states (two doubly degenerate levels at zero field) of Co$^{2+}$. In each circular polarization there are 16 possible transitions but only 4 are optically allowed due to requirement of Co$^{2+}$ spin conservation for electric dipole transition (solid arrows on Fig. \ref{NaprezeneCo_TransitionScheme}). In zero magnetic field the 4 lines for  $\sigma^+$ polarization coincide with those for $\sigma^-$.

Intensity of two inner and two outer lines reflects differences in occupancy of various Co$^{2+}$ states. Fig.~\ref{Fig4lines} shows a typical case, when cobalt has a fundamental state with spin projection $\pm3/2$ and consequently outer lines related to such a spin states are more pronounced than lines related to $\pm1/2$ states (inner lines). However we argue that apart from Co$^{2+}$ level ordering, such four main lines do not carry sufficient information about absolute value of Co$^{2+}$ zero-field splitting. It is because line separation energy is determined by $s,p$-$d$ exchange interaction not by splitting of Co$^{2+}$ states. Moreover, occupancy of various states of Co$^{2+}$ under optical excitation is not simply governed by temperature and Boltzmann distribution. One could try to describe it by an effective temperature, but there is no good method of determination of such a parameter, which could be order of magnitude larger than actual experimental temperature (e.g. 10 or 30 K instead of 1.7 K in Supplementary Information of Ref. \onlinecite{kobak-nature-2014}). Moreover, Fig. \ref{Fig4lines} shows that two outer lines of a QD with a single Co$^{2+}$ ion exhibit not-equal intensity which is a fingerprint of efficient relaxation of Co$^{2+}$ - exciton system during lifetime of a single exciton introduced to the QD. This makes measurement of occupancy distribution of Co$^{2+}$ states in empty dot even more complex. Therefore we conclude that precise determination of Co$^{2+}$ zero-field splitting from analysis of four main photoemission lines of a QD with a single Co$^{2+}$ is not feasible.

\begin{figure*}
\includegraphics[width=1\linewidth]{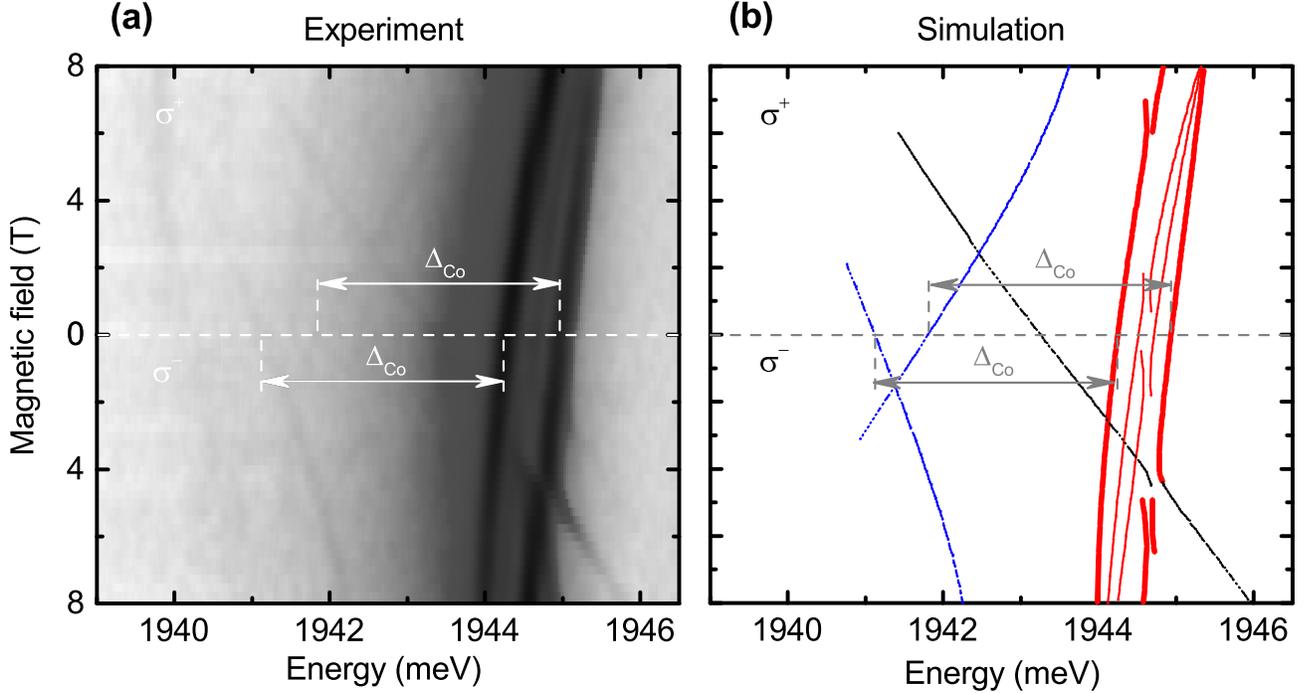}
\caption{(Color online) (a) Polarization resolved PL intensity map of neutral exciton in a CdTe/ZnTe QD with a single Co$^{2+}$ as a function of magnetic field in Faraday configuration, at T = 1.5 K. (b) Corresponding simulation of the optical transitions with the model described in the text. There are four main emission lines (red lines on panel (b)) corresponding to the bright exciton transition with conserved spin of Co$^{2+}$ and three weak emission lines: one associated with the dark exciton recombination (black line on panel (b)) which does not carry information about zero-field splitting and two lines related to the transitions allowed by cobalt $\pm3/2$ and $\mp1/2$ spin states mixing (blue lines on panel~(b)). As explained in Fig.~\ref{NaprezeneCo_TransitionScheme}, zero-field splitting of cobalt ($\Delta_{Co}=3.13$~meV) is determined by subtraction of energies for pairs of transitions for which initial state is the same ($\pm3/2$) but final state is different ($\pm3/2$ and $\mp1/2$). Values of $\Delta_{Co}$ parameter estimated from two pairs of emission lines are the same. }
\label{NaprezeneCo_KropkaWiekszeD}
\end{figure*}

\subsection{Partially allowed transitions and determination of Co$^{2+}$ zero-field splitting}

In previous section we discussed four main optical transitions with conserved spin of Co$^{2+}$ ion. In this section we report about additional bright neutral exciton transitions, which can be observed only if in--plane strain mixes Co$^{2+}$ spin states. Since such mixing is influenced by $s$,$p$-$d$ exchange interaction between carriers and magnetic ion, Co$^{2+}$ spin mixing is slightly different for a QD with and without exciton. When spin wavefunction of Co$^{2+}$ is not identical in initial and final state of optical transition, part of oscillator strength is distributed between transitions where main spin component is different in initial and final state, but both states are not fully orthogonal. This effect opens possibility of studying partially allowed optical transitions shown by blue and green dashed arrows in Fig.~\ref{NaprezeneCo_TransitionScheme}. The intensity of such emission lines is strongly sensitive on parameters describing Co$^{2+}$ ion. For values of parameters determined in this work, intensity ratio between main emission lines (red colour in Fig.~\ref{NaprezeneCo_TransitionScheme}) and partially allowed lines (blue and green color in Fig.~\ref{NaprezeneCo_TransitionScheme}) is 3 orders of magnitude. Measurements of such weak PL lines is experimentally challenging but it is worth of effort because partially allowed lines carry information about parameters describing spin structure of Co$^{2+}$ ion. In particular, energy difference between the strongest emission lines in spectra (red outer lines) and weak, partially allowed lines at lower energy (blue lines) are equal exactly to cobalt spin state splitting ($\Delta_{Co}$) since such transitions have the same initial states ($\pm3/2$ with admixture of $\mp1/2$) but different final states ($\pm3/2$ with admixture of $\mp1/2$ and $\pm1/2$ with admixture of $\mp3/2$, respectively). The same energy distance ($\Delta_{Co}$) is between inner main lines (red color) and partially allowed emission lines at higher energy (green color). The energy splitting of low energy weak emission lines (blue) is related to $s$,$p$-$d$ exchange interaction in initial state, so it is the same as splitting of outer main emission lines (red). Similarly, high energy weak emission lines (green) exhibit splitting equal to splitting of inner main lines (red).

Experimental observation of partially allowed transitions is shown in Fig.~\ref{NaprezeneCo_KropkaWiekszeD}(a) which presents typical low temperature magnetophotoluminescence of bright exciton in a QD with a single Co$^{2+}$ ion in Faraday configuration, measured in two circular polarizations of detection. Line identification on the spectra is in agreement with simulation shown in Fig.~\ref{NaprezeneCo_KropkaWiekszeD}(b). In both figures (a) and (b) we observe strong main emission lines (red lines on panel (b) or scheme of Fig. \ref{NaprezeneCo_TransitionScheme}) corresponding to transitions with conserved spin of Co$^{2+}$. Additionally we observe also three weak emission lines. The first  one is associated with the dark exciton recombination (black line on panel (b)) which involves change of Co$^{2+}$ spin from mainly +3/2 to mainly -3/2, so it exhibit the largest g-factor and it is a fingerprint of Co$^{2+}$ spin states mixing, but it does not carry information about zero-field splitting of Co$^{2+}$. Two more weak lines are related to bright exciton transitions partially allowed by cobalt $\pm3/2$ and $\mp1/2$ spin states mixing (blue~lines on panel (b) and scheme of Fig. \ref{NaprezeneCo_TransitionScheme}). We determine zero-field splitting of cobalt ($\Delta_{Co}=3.13$~meV for Co$^{2+}$ shown in~Fig.~\ref{NaprezeneCo_KropkaWiekszeD}) by subtraction of emission energies of lines related to the the transitions for which initial state is the same (mainly $\pm3/2$) but final state is different (mainly $\pm3/2$ and $\mp1/2$ for weak and strong lines respectively). Note that values of $\Delta_{Co}$ parameter estimated from two pairs of emission lines are the same.

\begin{figure*}
\includegraphics[width=1\linewidth]{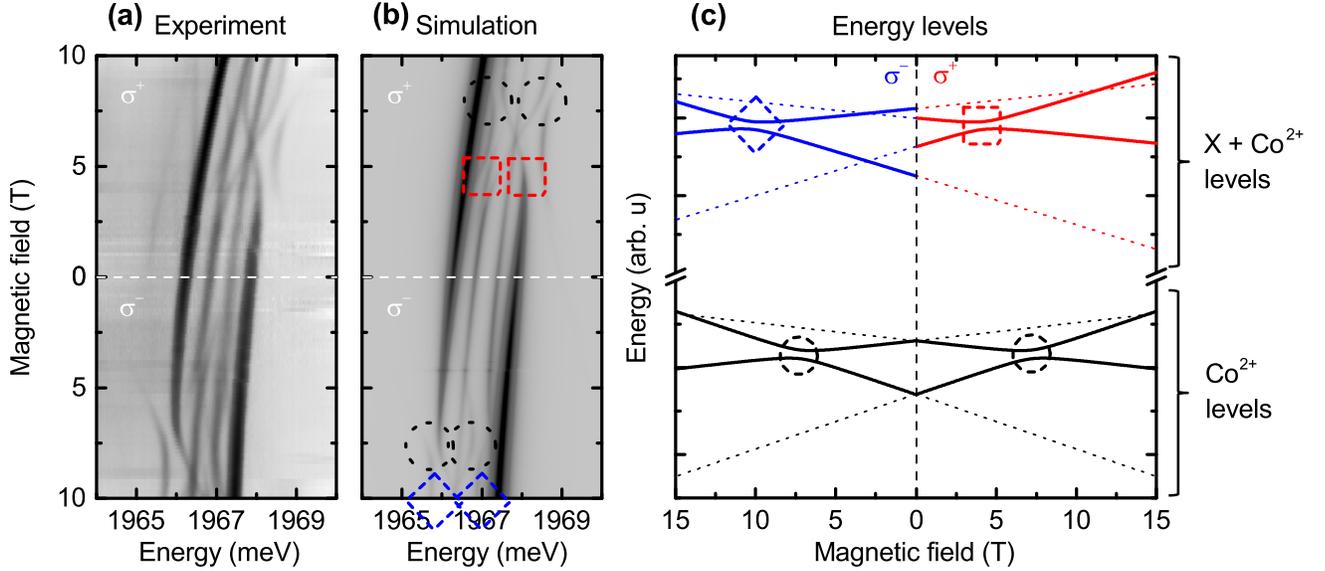}
\caption{
(Color online) (a) Polarization resolved magneto-PL intensity map in Faraday configuration at T = 1.5 K for a CdTe/ZnTe QD with a single Co$^{2+}$ ion exhibiting within experimental range an anticrossing of 3/2 and -1/2 spin states. (b) Corresponding simulation of the optical spectra. (c) Corresponding simulation of energies of initial state (exciton + Co$^{2+}$ system) and final state (empty dot with a single Co$^{2+}$) of optical transition. Anticrossing between $+3/2$ and $-1/2$ cobalt spin states in the absence of exciton appears four times on the spectra, at magnetic field of about 7$.5$~T and is denoted on panel (b) and (c) by black circles. Red and blue squares on panel (b) and (c) represent an anticrossings of cobalt spin states modified by the presence of neutral exciton which acts on the Co$^{2+}$ as effective magnetic field. As a consequence such anticrossings are shifted to about $4.5$~T and $10.5$~T for $\sigma^{+}$ and $\sigma^{-}$ polarization respectively.}
\label{NaprezeneCo_KropkaAleksandra}
\end{figure*}

\begin{figure}[!h]
\includegraphics[width=0.9\linewidth]{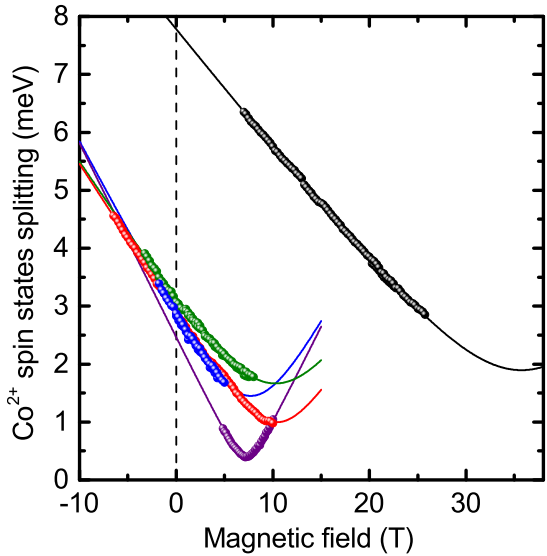}
\caption{(Color online) Points - energy difference between mixed $\pm3/2$ and $\mp1/2$ spin states of various Co$^{2+}$ ions in CdTe/ZnTe QDs, determined from PL spectra as a function of magnetic field, near the anticrossing. Solid lines - fit of Co$^{2+}$ states splitting given by Eq. \ref{eq:CoSplittingVsB} with fitting parameters summarized in Table~\ref{tab_parametersTable}. }
\label{NaprezeneCo_Co_splitting_vs_b}
\end{figure}

Simulation presented in Fig.~\ref{NaprezeneCo_KropkaWiekszeD} (b) is based on simple model of exciton - Co$^{2+}$ - ion system described in details in Methods of Ref. \onlinecite{kobak-nature-2014}. The most important part of our theoretical description is given by the standard Hamiltonian of Co$^{2+}$ ion, which is also the Hamiltonian of the final state of the system after the exciton recombination:
\begin{equation}
 \label{eq:CoHamiltonian}
\mathcal{H}_{Co}=g_{Co}\mu_B \overrightarrow{\vecb{B}}\cdot\overrightarrow{\vecb{J}} + D\vecb{J^{2}_{z}}+ E(\vecb{J^{2}_{y}}-\vecb{J^{2}_{x}}),
\end {equation}
where the first term represents the Zeeman effect described by magnetic ion \mbox{$g$-factor}, Bohr magneton $\mu_B$, magnetic field $\overrightarrow{\vecb{B}}$ and total magnetic momentum $\overrightarrow{\vecb{J}}$ ($J=3/2$). The second and the third term describe magnetic anisotropy of a magnetic impurity with $D$ and $E$ representing axial and in-plane components of crystal field. Both parameters $D$ and $E$ can be determined basing on anticrossings discussed in detail in the next section.

\subsection{Anticrossing of Co$^{2+}$ ion spin states and determination of crystal field parameters}

Since g-factors are different for main and partially allowed transitions (see Fig.~\ref{NaprezeneCo_KropkaWiekszeD}),  approaching of various lines in magnetic field is inevitable. Series of anticrossigs between main and partially allowed transitions are observed in magnetophotoluminescence experiment shown in Fig.~\ref{NaprezeneCo_KropkaAleksandra}(a). Additionally, Fig.~\ref{NaprezeneCo_KropkaAleksandra}(b) presents corresponding  simulation of the optical spectra and Fig.~\ref{NaprezeneCo_KropkaAleksandra}(c) presents simulated energy levels of neutral exciton in a CdTe QD with Co$^{2+}$ (initial states of photoluminescence) and empty dot with single cobalt (final states). All panels of  Fig.~\ref{NaprezeneCo_KropkaAleksandra} show important spectral features closely related to the spin structure of Co$^{2+}$ ion. Anticrossing between $+3/2$ and $-1/2$ cobalt spin states in the absence of exciton appears four times in excitonic spectra (due to final state of recombination), all appearances are at the same magnetic field (about 7$.5$~T for QD from Fig. \ref{NaprezeneCo_KropkaAleksandra}a). They are denoted on panel (b) and (c) by black circles. Red and blue squares on panel (b) and (c) represent anticrossings of cobalt spin states modified by the presence of carriers forming neutral exciton, which acts on Co$^{2+}$ as effective magnetic field of about 3 T for QD from Fig. \ref{NaprezeneCo_KropkaAleksandra}a. As a consequence such anticrossings are shifted and are visible at about $4.5$~T and $10.5$~T for $\sigma^{+}$ and $\sigma^{-}$ polarization of detection respectively. Observed shift is analogical to the shift of the magnetic ion levels anticrossing from zero field to field of about 2 T which has been already reported for Mn$^{2+}$+h complex in InAs/GaAs QDs,\cite{Kudelski_2007_PRL,Krebs_2009_PRB, Baudin_2011_PRL} and for Fe$^{2+}$ in CdSe/ZnSe QDs\cite{Smolenski_2016_Nature_Comm} where $s$,$p$-$d$ exchange interaction with exciton was also acting on magnetic ion as an effective magnetic field.

Particularly important for this work are anticrossings in the final state which are marked in Fig.~\ref{NaprezeneCo_KropkaAleksandra} by black circles. They give additional information about Co$^{2+}$ properties in absence of carriers. In such an anticrossing separation energy of spectral lines corresponds directly to separation energy between Co$^{2+}$ states close to the anticrossing of spin states \mbox{+3/2} and~\mbox{-1/2}. In Fig.~\ref{NaprezeneCo_Co_splitting_vs_b} we plot the dependence of such Co$^{2+}$ spin states splitting $\Delta_{Co}(B)$ as a function of the magnetic field for QDs shown in Fig. \ref{NaprezeneCo_KropkaWiekszeD} (violet points), Fig.~\ref{NaprezeneCo_KropkaAleksandra} (green points), and for three other QDs with a single Co$^{2+}$ ion.

Solving of Hamiltonian (\ref{eq:CoHamiltonian}) leads us to analytical formula describing  magnetic field evolution of energy distance between mixed $\pm3/2$ and $\mp1/2$ spin states of Co$^{2+}$:
\begin{equation} 
\label{eq:CoSplittingVsB}
\Delta_{Co}(B)=2\sqrt{3E^2+(g_{ion}\mu_B B+D)^2}
\end{equation}
where parameters $D$ and $E$ (axial and in-plane components of crystal field) can be extracted directly from the experimental data. Parameter $E$ can be determined from the value of $\Delta_{Co}$ for magnetic field at which anticrossing between $+3/2$ and $-1/2$ spin states appears \mbox{($\Delta_{Co}(B_{min\Delta})=2\sqrt{3}E$)}. Knowing $E$ parameter one can obtain $D$ parameter from the zero-field spin states splitting given by formula: \mbox{$\Delta_{Co}(0)=2\sqrt{3E^2+D^2}$}. Finally \mbox{$g$-factor} is equal to $-D/(\mu_BB_{min\Delta})$. Parameters $D$, $E$, and the \mbox{$g$-factor} of Co$^{2+}$ ion can be also determined by fit of Eq. \ref{eq:CoSplittingVsB} to experimental data, as shown by solid lines in Fig.~\ref{NaprezeneCo_Co_splitting_vs_b}.

\begin{table}[b]
\begin{tabular}{|c|c|c|c|}
\hline
$~\Delta_{Co}(0)~$ & $D$ (meV) & $E$ (meV) & $g_{Co}$ \\
(meV) &  &   &   \\
\hline
$2.47$ & $-1.22 \pm 0.01$ & $0.12 \pm 0.01$ & $2.91 \pm 0.02$ \\
$2.86$ & $-1.23 \pm 0.02$ & $0.42 \pm 0.02$ & $2.76 \pm 0.09$ \\
$2.91$ & $-1.37 \pm 0.01$ & $0.29 \pm 0.01$ & $2.27 \pm 0.01$ \\
$3.13$ & ~~$-1.33 \pm 0.01$~~ & ~~$0.48 \pm 0.01$ ~~& ~ $2.23 \pm 0.03$~~ \\
$7.78$ & $-3.77 \pm 0.01$ & $0.55 \pm 0.02$ & $1.82 \pm 0.02$ \\
\hline
\end{tabular}
\caption{Values of key parameters describing spin configuration and anisotropy of Co$^{2+}$ ion for all studied QDs: zero-field splitting $\Delta_{Co}(0)$, axial ($D$) and in-plane ($E$) components of crystal field and $g_{Co}$. Measurement of the largest zero field splitting (the last line) required high magnetic field laboratory in Grenoble.}
\label{tab_parametersTable}
\end{table}

The values of the parameters describing Co$^{2+}$ ions in all studied QDs are summarized in Table~\ref{tab_parametersTable}. The data are sorted in increasing order with respect to the zero-field splitting $\Delta_{Co}$, which varies from about $2.5$~meV to almost $8$~meV. The value of parameter $D$ is in the range between $-1.22$~meV and $-3.77$~meV. For all discussed QDs parameter $D$ is negative which favours states of spin equal $\pm3/2$. Parameter $E$, which causes the mixing of Co$^{2+}$ spin states, varies from $0.12$ meV to $0.55$ meV. The values of Co$^{2+}$ \mbox{$g$-factor} strongly differ from each other starting from $1.82$ and ending with the value $2.91$. 
Analysis of the Table~\ref{tab_parametersTable} shows that there is an anti-correlation between $\Delta_{Co}$ and $g_{Co}$. The smaller Co$^{2+}$ zero-field splitting the larger value of Co$^{2+}$ \mbox{$g$-factor}. Surprisingly, for the declining value of Co$^{2+}$ zero-field splitting we do not observe the convergence of Land\'e factor of Co$^{2+}$ to the limit defined by unstrained CdTe crystal,\cite{Ham_1960_PRL} which is equal $2.31$. This shows that there are other important factors affecting the structure of the energy states of cobalt ion. One of possible explanations of g-facotor lower than 2.31 is that larger g-factor would be observed for other axis than growth axis tested in our experiment.

We note that distribution of measured values of zero-field splitting is probably not governed by distribution in the studied samples, but it is rather affected by preselection of studied Co$^{2+}$ ions caused by our experimental limitations. At zero field or at magnetic field in range of a few Tesla, only relatively small zero-field splitting of Co$^{2+}$ can be determined. For large zero-field splitting partially allowed transitions are too weak to be observed. Therefore we expect that real distribution Co$^{2+}$ zero-field splitting in the samples is even wider than observed and that typical value is larger than about 3 meV observed a few times in our first experiments. This expectation is supported by high magnetic field experiment (black data of Fig. \ref{NaprezeneCo_Co_splitting_vs_b}), where anticrossing of +3/2 and -1/2 is expected at about 35~T for Co$^{2+}$ ion with $\Delta_{Co}(0)$ equal to almost 8~meV and parameter $D=-3.77$~meV.

Absolute values of parameter $D$ obtained in this work are significantly larger than values reported for semiconductor systems without strain. For Co$^{2+}$ in zinc blende semiconductors like CdTe, ZnTe or ZnSe parameter $D$ is neglected\cite{Ham_1960_PRL,Wu_epr_2004,Grzybowski_SSC_2015}. In wurtzite structure diluted magnetic semicondutors with Co$^{2+}$ parameter $D$ was found to be positive with absolute value much smaller than in our experminets on QDs: +0.06~meV for CdSe,\cite{Lewicki_PRB_1991,Isber_PRB_1995} +0.08~meV for CdS,\cite{Lewicki_PRB_1991,Bindilatti_PRB_1994} and +0.34~meV for ZnO.\cite{Estle_1961_BAPS,Macfarlane_PRB_1970,Koidl_1977_PRB,Jedrecy_PRB_2004,Ferrand_2005_JS,Sati_PRL_2006,Pacuski_PRB_2006}

In strained semiconductor systems, parameter $D$ obtained in this work for Co$^{2+}$ can be compared to values obtained for transition metal ions. For Mn$^{2+}$ QDs CdTe/ZnTe, parameter $D \approx 0.007$~meV was determined from depolarization efficiency of optically oriented ion\cite{LeGall_2009_PRL} and  from variation of coherent oscillation amplitude,\cite{Goryca_PRL_2014} which were also used for determination of parameter $E = 1.8$~$\mu$eV.\cite{Lafuente_Sampietro_2016_PRB} Much larger values, in range of~meV, are expected from magneto-photoluminescence of main emission lines in QDs with transition metal ions exhibiting non-zero orbital momentum, $D>0.8$~meV\cite{Smolenski_2016_Nature_Comm} for Fe$^{2+}$,  $D>2.25$~meV\cite{Lafuente_Sampietro_2016_PRB} for Cr$^{2+}$.
Finally, in our first report on main emission lines of a QD CdTe/ZnTe with a single Co$^{2+}$, using multi parameter fit we obtained value $D_z \approx -1.4$~meV,\cite{kobak-nature-2014} in the same range as directly determined values presented in this work.

\section{Conclusions}
We presented optical studies of individual Co$^{2+}$ ions in epitaxial QDs. Using partially allowed optical transitions we describe strain induced zero-field splitting of Co$^{2+}$ ions and anticrossing of Co$^{2+}$ spin states. From this anticrossing we determine Hamiltonian parameters $D$ and $E$ representing axial and in-plane components of crystal field. Our technique, can be applied for optical measurements of zero-field splitting of other transition metal ions (e.g. Cr$^{2+}$, Fe$^{2+}$) in QDs, providing that very low noise spectra or high enough magnetic fields are available.

\begin{acknowledgments} This work was partially supported by the Polish National Science Centre under decision DEC-2015/18/E/ST3/00559, DEC-2011/02/A/ST3/00131, DEC-2013/09/B/ST3/02603, DEC-2012/05/N/ST3/03209, by Polish Ministry of Science and Higher Education programme „Iuventus Plus” in years 2015-2017 project number IP2014 034573, and in years 2013$-$2017 as research grant "Diamentowy Grant". The two of us (JK and TS) were supported by the Foundation for Polish Science through the START programme. Project was carried out with the use of CePT, CeZaMat, and NLTK infrastructures financed by the European Union - the European Regional Development Fund within the Operational Programme "Innovative economy".
\end{acknowledgments}

\end{document}